\documentclass[12pt]{article}

\newcommand{\be}{\begin{equation}}
\newcommand{\ee}{\end{equation}}
\newcommand{\bi}[1]{\vspace{-3mm} \bibitem{#1}}

\usepackage{epsfig,amsmath,amssymb,graphics,graphicx} 

\usepackage{amscd}
\usepackage{pb-diagram}
\usepackage{hyperref}
\usepackage{amsfonts}

\begin{document}

\begin{center}

{\it Physica A: Statistical Mechanics and its Applications.
Vol.421. (2015) 330-342.}
\vskip 3mm

{\bf \large Fractional Liouville Equation on Lattice Phase-Space} \\

\vskip 7mm
{\bf \large Vasily E. Tarasov} \\
\vskip 3mm

{\it Skobeltsyn Institute of Nuclear Physics,\\ 
Lomonosov Moscow State University, \\
Moscow 119991, Russia} \\
{E-mail: tarasov@theory.sinp.msu.ru} \\

\begin{abstract}
In this paper we propose 
a lattice analog of phase-space fractional Liouville equation. 
The Liouville equation for phase-space lattice with
long-range jumps of power-law types is suggested.
We prove that the continuum limit transforms
this lattice equation into Liouville equation
with conjugate Riesz fractional derivatives
of non-integer orders with respect to coordinates of
continuum phase-space.
An application of the fractional Liouville equation with 
these Riesz fractional derivatives to describe
properties of plasma-like nonlocal media is considered.
\end{abstract}

\end{center}



\section{Introduction}

A consistent formulation of the nonlocal statistical mechanics
was first constructed by Vlasov in \cite{Vlasov1,Vlasov2}.
Fractional calculus 
\cite{SKM,Podlub,KST,Uch,Mainardi1997,FC1,FC2}
has a lot of applications in physics 
\cite{SATM,LA,Mainardi,TarasovSpringer,KLM,IJMPB2013,US}
and it allows us to take into account
fractional power-law nonlocality of 
continuously distributed systems.
Using the fractional calculus, we can consider
fractional differential equations for
conservation of probability in generalized phase spaces.
The Liouville equation with phase-space fractional derivatives
allows us to formulate fractional statistical mechanics
that describes systems with power-law nonlocality. 
The fractional statistical mechanics can be considered
as a special form of the nonlocal statistical mechanics 
\cite{TarasovSpringer}.
It should be noted that the use of the fractional derivative 
of non-integer order is actually equivalent to using 
an infinite number of derivatives of integer orders,
which can be arbitrarily large values 
(see Lemma 15.3 in \cite{SKM}).

We note that fractional Liouville and Bogoliubov equations
are discussed in 
\cite{TFLE-1,TFLE-2,TFLE-3,TFLE-4,TFLE-5,TFLE-6,TFLE-7}.
There are two different types of approaches 
to space-fractional generalization of the Liouville equation:

1) The first approach to  
generalization of the Liouville equation in the framework of
fractional dynamics has been suggested in 
\cite{TFLE-1,TFLE-2,TFLE-3,TFLE-4}.
This generalization is derived from the normalization condition
with the fractional integration 
over phase-space coordinates.
This fractional normalization condition 
is interpreted as an equation in 
fractional-dimensional phase-space 
or in phase-space with fractional powers 
of phase-space coordinates.
We should note that the fractional Liouville equations,
which are suggested in \cite{TFLE-1,TFLE-2,TFLE-3,TFLE-4},
do not contain fractional derivatives of non-integer orders.

2) The second approach, which is based on the Liouville equation 
with fractional-order derivatives, has been suggested in 
\cite{TFLE-5,TFLE-6,TarasovSpringer}, where 
the Caputo type of fractional derivatives is used.
To obtain this equation, we use the conservation of
probability for a fractional differential
volume element of the phase space. 
The suggested fractional Liouville equation 
is used to derive the fractional kinetics equations.
The Bogoliubov hierarchy equations with fractional derivatives 
with respect to phase space coordinates 
are derived in \cite{TFLE-5,TFLE-6}. 
The Vlasov equation and the Fokker-Planck equation
with the Caputo fractional derivatives are also obtained 
from the fractional Liouville equations.
In \cite{TFLE-7} the Liouville equation 
with the Caputo fractional derivatives is used 
to describe media with spatial dispersion of power-law type 
that is considered in \cite{AP2013}.
We should note that this phase-space fractional 
Liouville equation are considered 
only for Caputo fractional derivatives.

As it was shown in \cite{JPA2006,JMP2006,JPA2014} 
(see also \cite{Chaos2006,CNSNS2006,LZ2006,IS2012} and
\cite{CEJP2013,MOM2014,ISRN-CMP2014,IJSS2014,ND2014,AHEP2014,JSM2014}), 
the continuum equations with fractional derivatives 
can be directly connected to
lattice models with long-range properties. 
Long-range interaction and properties are important
for different problems in statistical mechanics 
\cite{LLI-1,LLI-2,LLI-3},
in kinetic theory and non-equilibrium statistical mechanics 
\cite{LLI-4,LLI-5},
in the theory of non-equilibrium phase transitions
\cite{LLI-6,LLI-7}.
A connection between the dynamics of lattice system 
with long-range properties 
and the fractional continuum equations is proved 
by using the special transform operation 
\cite{JPA2006,JMP2006,JPA2014}
and it has been applied to different subjects
\cite{CEJP2013,MOM2014,ISRN-CMP2014,IJSS2014,ND2014,AHEP2014,JSM2014}).

In this paper, we propose Liouville equation 
for lattice phase-space.
This Liouville equation describes 
fractional dynamics of a distribution function
on unbounded homogeneous lattice with long-range jumps 
from one site to other sites.
We prove that continuous limit of
the suggested lattice Liouville equation gives 
the fractional Liouville equation 
for continuum phase-space.
This fractional Liouville equation contains
generalized conjugate Riesz derivatives 
of non-integer orders with respect to
coordinates and momenta.
As an example, we consider an application
of the fractional Liouville equation with 
these Riesz derivatives 
to describe properties of plasma-like nonlocal media.


\section{Liouville equation for lattice with long-range jumps}

In statistical mechanics \cite{Liboff}, 
the basic concept is the ensemble 
that is a set of classical systems identical 
in nature, which are subjected to forces determined 
by identical laws \cite{Gibbs}, 
but distributed in phase-space.
We can state that the statistical ensemble is a set of 
independent systems identical in equations of motion, 
but differing in their initial conditions \cite{Ch16}.
It is usually assumed that the systems are 
continuously distributed in the phase-space. 
In this paper, we consider systems of particles
that are distributed in lattice phase-space.
Note that the statistical method should be used not only 
for systems with a very large number of particles. 
It is necessary to use this method in every case, even 
that of a single particle in 
the simplest possible conditions \cite{Born}.
In the Liouvillian picture, we can describe 
the dynamics of a statistical ensemble 
in the phase-space points (sites)
through which the ensemble points move. 
In this case, properties of the ensemble 
are assigned to points (sites) in phase-space at each given time, 
without an attempt to identify the individual system of the ensemble.
In this paper, we consider a statistical ensemble 
on the lattice phase-space.


Let us consider a system of classical identical $N$ spinless 
point particles, 
which are characterized by the phase-space coordinates 
$q_{s,k}$ and $p_{s,k}$ with $s=1, \ldots , N$, $k=1,2,3$, 
As a lattice analog of continuum phase-space for this system 
we will use an unbounded lattice 
in $6N$-dimensional Euclidean phase-space $\mathbb{R}^{6N}$.
The lattice is characterized by space periodicity.
For unbounded phase-space lattice 
we define $6N$ non-coplanar vectors 
${\bf a}_1$, \ldots , ${\bf a}_{6N}$, that 
are the shortest vectors by which a lattice can 
be displaced such that it is brought back into itself. 
For simplification, we consider a lattice with mutually 
perpendicular lattice vectors ${\bf a}_j$.
We choose the Cartesian coordinate system 
such that the directions of the axes  
coincide with the vector ${\bf a}_j$, such that 
${\bf e}_j = {\bf a}_j/|{\bf a}_j|$ 
are the basis vectors of the Cartesian coordinate system
in $\mathbb{R}^{6N}$.
This simplification means that we consider 
$6N$-dimensional analog of 
the primitive orthorhombic Bravais lattice.

Choosing the coordinate origin at one of the sites, 
all phase-space lattice sites can be numbered by 
the vectors ${\bf n}_q = (n_1, \ldots ,n_{3N})$
and ${\bf n}_p = (n_{3N+1}, \ldots ,n_{6N})$,
where $n_j$ ($j=1,\ldots,6N$) are integer. 
The position of a lattice site is defined by the vector
\be
{\bf x}({\bf n})= {\bf x}({\bf n}_q) +{\bf x}({\bf n}_p) =
\sum^{6N}_{j=1} n_j{\bf a}_j , \quad
\ee
where
\be
{\bf x}({\bf n}_q) = \sum^{3N}_{j=1} n_j{\bf a}_j , \quad
{\bf x}({\bf n}_p) = \sum^{6N}_{j=3N+1} n_j{\bf a}_j .
\ee

Considering the statistical ensemble on 
the phase-pace lattice, the $N$-particle system
is represented by a point-particle, 
which is moved through the lattice sites.
This lattice particle, which is lattice analog 
of the $N$-particle system, will be called 
the lattice quasi-particle.
We assume that it can be localized in the sites 
of the phase-space lattice, i.e. 
the positions of the quasi-particle 
coincide with the lattice sites.
Then the vectors ${\bf n}_q$ and ${\bf n}_p$ 
can be used to describe the quasi-particle. 
We will consider long-range jumps of 
the quasi-particle on the lattice phase-space.

The distribution function of lattice quasi-particle, 
which is the probability density 
to be in the lattice site, 
will be denoted by 
$\rho ({\bf n}_q,{\bf n}_p,t) = \rho (n_1, \ldots ,n_{6N},t)$,
where the vectors ${\bf n}_q$ and ${\bf n}_p$ define the site.
We can use the notation $\rho_L ({\bf n}_q,{\bf n}_p,t)$
instead of $\rho ({\bf n}_q,{\bf n}_p,t)$
to emphasize that it is a lattice function.
The function $\rho ({\bf n}_q,{\bf n}_p,t)$ 
satisfies the conditions
\be
\sum^{+\infty}_{ n_1 = - \infty } \ldots 
\sum^{+\infty}_{ n_{6N} = -\infty } 
\rho (n_1, \ldots , n_{6N},t) = 1 ,
\quad \rho (n_1, \ldots ,n_{6N},t) \ge 0 
\quad (t \in \mathbb{R}).
\ee
In the lattice, any particle of the $N$-particle system 
is described by the six numbers
\be
{\bf n}(s) =
(n_{s},n_{s+1},n_{s+2},n_{3N+s},n_{3N+s+1},n_{3N+s+2}) ,
\quad (s=1, \ldots , N) .
\ee
Since we assume that the system consists of 
the identical $N$ particles, 
then the function $\rho (n_1, \ldots , n_{6N},t)$ 
has the symmetry with respect to permutations
$ {\bf n}(s) \ \longleftrightarrow \ {\bf n}(s^{\prime})$,
where $1 \le s, s^{\prime} \le N$.
We also can define reduced $s$-particle distribution 
functions by the equation
\be
\rho_s ({\bf n}_q,{\bf n}_p,t) =
\sum_{{\bf n}(s+1)} \ldots \sum_{{\bf n}(N)} 
\rho ({\bf n}_q,{\bf n}_p,t) ,
\ee
where $\rho_s ({\bf n}_q,{\bf n}_p,t)$
depends on the vectors ${\bf n}(1), \ldots {\bf n}(s)$ only.
In the case $s=N$, we can see that 
$\rho_N ({\bf n}_q,{\bf n}_p,t)=\rho ({\bf n}_q,{\bf n}_p,t)$.

Let us consider the equation for distribution function  
on unbounded homogeneous phase-space lattice in the form
\[
\frac{\partial \rho ({\bf n}_q,{\bf n}_p,t) }{\partial t} 
+ \sum^{3N}_{j=1} 
\mathbb{D}^{-}_{L,q} \left[ \alpha_j \atop j \right]
\, \Bigl( V_j ({\bf m}_q,{\bf m}_p,t) 
\, \rho ({\bf m}_q,{\bf m}_p,t) \Bigr) +
\]
\be \label{FLE-L}
+ \sum^{6N}_{j=3N+1} 
\mathbb{D}^{-}_{L,p} \left[ \beta_j \atop j \right]
\, \Bigl( F_j ({\bf m}_q,{\bf m}_p,t) \,
\rho ({\bf m}_q,{\bf m}_p,t) \Bigr) =
J ({\bf n}_q,{\bf n}_p,t) ,
\ee
where the lattice function $\rho ({\bf n}_q,{\bf n}_p,t)$ 
is the probability density 
to find the quasi-particle at site $({\bf n}_q,{\bf n}_p)$ 
at time $t$.
In equation (\ref{FLE-L}), the term 
$J ({\bf n}_q,{\bf n}_p,t)$ describes an external source.
The lattice operators 
$\mathbb{D}^{-}_{L,q} \left[ \alpha_j \atop j \right]$
and $\mathbb{D}^{-}_{L,p} \left[ \beta_j \atop j \right]$
are defined by 
\be \label{Dalpha1}
\mathbb{D}^{-}_L \left[ \alpha_j \atop j \right] 
f ({\bf m}_q,{\bf m}_p,t) = 
\frac{1}{a^{\alpha_j}_j} \sum_{m_j=-\infty}^{+\infty} \; 
K^{-}_{\alpha_j} (n_j-m_j) \; f ({\bf m}_q,{\bf m}_p,t) ,
\ee
where we have 
$\mathbb{D}^{-}_L \left[ \alpha_j \atop j \right]=
\mathbb{D}^{-}_{L,q} \left[ \alpha_j \atop j \right]$
for $j=1, \ldots, 3N$ and
$\mathbb{D}^{-}_L \left[ \alpha_j \atop j \right]=
\mathbb{D}^{-}_{L,p} \left[ \beta_j \atop j \right]$
with $\beta_j=\alpha_j$ for $j=3N+1, \ldots , 6N$.
The kernel $K^{-}_{\alpha_j} (n_j-m_j)$ 
of the operator (\ref{Dalpha1}) describes 
the quasi-particle long-range jumps 
with length $n_j - m_j$ on the phase-space lattice. 
If $K^{-}_{\alpha_j}(n_j-m_j)>0$ then the kernel 
characterizes the jump to the site with $n_j$ 
from all other sites with $m_j \ne n_j$. 
If $K^{-}_{\alpha_j}(n_j-m_j)<0$ then the kernel 
describes the reverse process.
In equation (\ref{FLE-L}), the parameters $\alpha_j$ 
(and $\beta_j$) are positive real numbers 
that characterize how quickly the intensity 
of the jumps in the lattice
decrease with increasing the jump length $|n_j - m_j|$.
These parameters also can be considered as degrees 
of the power-law spatial dispersion 
in the lattice \cite{CEJP2013,ISRN-CMP2014}.
Equation (\ref{FLE-L}) is the lattice 
fractional Liouville equation that 
describes long-range jumps 
on $6N$-dimensional phase-space lattice.

Let us consider the kernel $K^{-}_{\alpha_j} (n_j-m_j)$ 
in the form 
\be \label{Kn1-}
K^{-}_{\alpha_j}(n_j-m_j) = -\frac{\pi^{\alpha_j+1} 
\, (n_j-m_j)}{\alpha_j+2} \, _1F_2 
\left(\frac{\alpha_j+2}{2}; \frac{3}{2},\frac{\alpha_j+4}{2};
-\frac{\pi^2\, (n_j-m_j)^2}{4} \right) , \quad \alpha_j>-2,
\ee
where $\, _1F_2$ is the Gauss hypergeometric function 
\cite{Erdelyi}. 
Note that expression (\ref{Kn1-}) 
can be used not only for $\alpha_j>0$,
but also for some negative values of $-2<\alpha_j<0$.
In the kernel notation, the minus is used to mark that 
it is the odd functions of integer variable 
$n \in \mathbb{Z}$ such that
$K^{-}_{\alpha_j}(n_j-m_j) = - K^{-}_{\alpha_j}(m_j-n_j)$ 
for all $n_j, m_j \in \mathbb{Z}$ and $j=1, \ldots 6N$.

The discrete Fourier transform
\be \label{Jak-}
\hat{K}^{-}_{\alpha_j}(k_j) = \sum^{+\infty}_{n_j=-\infty} 
e^{-i k_j n_j} K^{-}_{\alpha_j}(n_j) = 
- 2 \, i \, \sum^{\infty}_{n_j=1} K^{-}_{\alpha_j}(n_j) 
\sin(k_jn_j) 
\ee
of the kernels $K^{-}_{\alpha_j}(n_j)$ 
defined by (\ref{Kn1-}) has the form
\be \label{AR-}
\hat{K}^{-}_{\alpha_j}(k_j)= 
i \, \operatorname{sgn}(k_j) \, |k_j|^{\alpha_j} .
\ee
Note that the function (\ref{Kn1-})
can obtained by the equation
\be \label{InverseK}
K^{-}_{\alpha_j}(n) = - \frac{1}{\pi} 
\int^{\pi}_0 k^{\alpha_j} \, \sin(n \, k) \, dk 
\ee
that is the inverse relation to equation (\ref{Jak-}). 
For integer values of $\alpha_j$, we can get 
simpler equations instead of (\ref{Kn1-}). 
For example, we have 
\be \label{1-3}
K^{-}_1(n) = \frac{(-1)^n}{n} , \quad
K^{-}_2(n) = \frac{(-1)^n\, \pi}{n} +
\frac{2(1-(-1)^n)}{\pi \, n^3}, \quad
K^{-}_3(n) = \frac{(-1)^n\, \pi^2}{n} - 
\frac{6 \, (-1)^n}{n^3} ,
\ee
where $(1-(-1)^n)=2$ for odd $n$, and 
$((-1)^n-1)=0$ for even $n$.

In the general case, we can consider the kernels 
\be \label{AR-ass}
\hat{K}^{-}_{\alpha_j}(k_j) = 
i \, \operatorname{sgn}(k_j) \, |k_j|^{\alpha_j} + 
o(|k_j|^{\alpha_j}), 
\quad ( k_j \rightarrow 0 ) ,
\ee
where the little-o notation $o(|k_j|^{\alpha_j})$ 
means the terms that include higher powers of 
$|k_j|$ than $|k_j|^{\alpha_j}$. 
The form (\ref{AR-ass}) means that we consider lattices 
with weak spatial dispersion \cite{CEJP2013}.
If we use condition (\ref{AR-ass}) instead of (\ref{AR-}),
then we can consider wider class of kernels to describe 
the long-range lattice jumps. 
As an example of the kernel with (\ref{AR-ass}), 
we give
\be \label{Kn2-}
K^{-}_{\alpha_j}(n) = \frac{ (-1)^{(n + 1)/2} \, 
\left( 2[(n+1)/2] -n \right) \, \Gamma (\alpha_j+1)}{
2^{\alpha_j} \, \Gamma((\alpha_j+n)/2 +1) \Gamma((\alpha_j-n)/2+1)} ,
\ee
where the brackets $[ \ ]$ mean the integral part, i.e.,
the floor function that maps a real number 
to the largest previous integer number.
The expression $\left( 2[(n+1)/2] -n \right)$ 
is equal to zero for even $n=2m$,
and it is equal to 1 for odd $n=2m-1$.
Note that the kernel (\ref{Kn2-}) is real valued function
since we have zero, when the expression $(-1)^{(n + 1)/2}$ 
becomes a complex number.
It is easy to see that we can use equation (\ref{Kn2-})
for all integer values $n \in \mathbb{Z}$.

For wide class of physical $N$-particles systems, we can use 
\be
V_j ({\bf m}_q,{\bf m}_p,t) = V_j ({\bf m}_p) , \quad
F_j ({\bf m}_q,{\bf m}_p,t) = F_j ({\bf m}_q,t) .
\ee
These conditions mean that $V_j$ are
components of the lattice analog of the particle velocity vector,
and $F_j$ are components of the lattice analog of 
the force vector that is independent of the particle momenta. 
In this case, the Liouville equation (\ref{FLE-L})
can be rewritten in the form
\[
\frac{\partial \rho ({\bf n}_q,{\bf n}_p,t) }{\partial t} 
+ \sum^{3N}_{j=1} V_j ({\bf m}_p) \,
\mathbb{D}^{-}_{L,q} \left[ \alpha_j \atop j \right]
\, \rho ({\bf m}_q,{\bf m}_p,t) +
\]
\be \label{FLE-L2}
+ \sum^{6N}_{j=3N+1} F_j ({\bf m}_q,t) \,
\mathbb{D}^{-}_{L,p} \left[ \beta_j \atop j \right]
\, \rho ({\bf m}_q,{\bf m}_p,t) =
J ({\bf n}_q,{\bf n}_p,t) .
\ee
In the general case, the usual Leibniz rule 
for the lattice fractional derivative
$\mathbb{D}^{-}_{L}$ does not hold if $\alpha_j \ne 1$.
This means that we have the inequality
\be
\mathbb{D}^{-}_{L,q} \left[ \alpha_j \atop j \right]
\, \Bigl( f ({\bf m}_q) \, g ({\bf m}_q) \Bigr) \ne
f ({\bf m}_q) \, 
\mathbb{D}^{-}_{L,q} \left[ \alpha_j \atop j \right]
\, g ({\bf m}_q) +
g ({\bf m}_q) \, 
\mathbb{D}^{-}_{L,q} \left[ \alpha_j \atop j \right]
\, f ({\bf m}_q) .
\ee
This property is analogous to the characteristic property
of derivatives of all fractional orders
and all integer orders $\alpha_j \ne 1$ \cite{CNSNS2013}.


\section{Fractional Liouville equation for 
phase-space continuum}

In this section, we use the methods suggested 
in \cite{JPA2006,JMP2006,JPA2014,TarasovSpringer}
to derive a fractional Liouville equation for 
phase-space continuum with power-law non-localities.

Let us define a transform operation 
that allows us to derive 
the fractional Liouville equation for continuum phase-space
from 
the Liouville equation for lattice phase-space (\ref{FLE-L}).\\

{\bf Definition}
{\it The lattice-continuum transform operation 
${\cal T}_{L \to C}$ is the combination 
\be \label{LST}
{\cal T}_{L \to C} =
{\cal F}^{-1} \circ \operatorname{Lim} \circ \ {\cal F}_{\Delta} 
\ee
of the following three operations: \\

1) The operation
{\bf ${\cal F}_{\Delta}$} is the discrete Fourier transform 
$f_L ({\bf n}_q,{\bf n}_p) \to {\cal F}_{\Delta}\{ f_L ({\bf n}_q,{\bf n}_p) \}
= \hat{f} ({\bf k}_q,{\bf k}_p) $ 
of the lattice function $f ({\bf n}_q,{\bf n}_p)$ 
that is defined by
\be \label{ukt}
\hat{f}({\bf k}_q,{\bf k}_p) 
= {\cal F}_{\Delta} \{ f_L ({\bf n}_q,{\bf n}_p) \} = 
\sum_{n_1=-\infty}^{+\infty} \ldots
\sum_{n_{6N}=-\infty}^{+\infty}
\, f_L (({\bf n}_q,{\bf n}_p) ) 
\; e^{-i \, ({\bf k}, {\bf x}({\bf n})) },
\ee
where
$({\bf k}, {\bf x}({\bf n})) =
({\bf k}_q, {\bf x}({\bf n}_q)) + 
({\bf k}_p, {\bf x}({\bf n}_p))$, 
\[
{\bf x}({\bf n}) = \sum^{6N}_{j=1} n_j \, {\bf a}_j , \quad
{\bf x}({\bf n}_q) = \sum^{3N}_{j=1} n_j \, {\bf a}_j , \quad
{\bf x}({\bf n}_p) = \sum^{6N}_{j=3N+1} n_j \, {\bf a}_j , \
\]
and $a_j=2\pi/k_{j0}$ is distance between lattice particle 
in the direction ${\bf a}_j$. 
Equation (\ref{ukt}) means that we consider the lattice function 
$f_L ({\bf n}_q,{\bf n}_p)$ as discrete Fourier coefficients
of some function $\hat{f}({\bf k}_q,{\bf k}_p)$ for 
$k_j \in [-k_{j0}/2, k_{j0}/2]$, where $j=1,\ldots,6N$. \\

2) The operation
{\bf $\operatorname{Lim}$} is
the passage to the limit $\hat{f} ({\bf k}_q,{\bf k}_p)
\to \operatorname{Lim} \{\hat{f} ({\bf k}_q,{\bf k}_p)\}
= \tilde{f} ({\bf k}_q,{\bf k}_p)$, where we use
$a_j \to 0$ (or $k_{j0} \to \infty$).
It allows us to derive the function 
$\tilde{f} ({\bf k}_q,{\bf k}_p)$
from $\hat{f} ({\bf k}_q,{\bf k}_p)$, where 
$\tilde{f} ({\bf k}_q,{\bf k}_p)$ is the Fourier integral transform
of the continuum function $f_C ({\bf q},{\bf p})$,
and $\hat{f}({\bf k}_q,{\bf k}_p)$ 
is the discrete Fourier transform 
of the lattice function 
$f_L ({\bf n}_q,{\bf n}_p)$, where 
\[
f_L ({\bf n}_q,{\bf n}_p)
= \left( \prod^{6N}_{j=1}(2 \pi / k_{j0}) \right)
f_C ({\bf x}({\bf n}_q),{\bf x}({\bf n}_p)), 
\]
and $n_j a_j= 2 \pi n_j /k_{j0} \to q_j$ for $j=1,\ldots,3N$
and 
$n_j a_j= 2 \pi n_j /k_{j0} \to p_j$ for $j=3N+1,\ldots,6N$. \\

3) The operation
{\bf ${\cal F}^{-1}$} is the inverse integral Fourier transform
$\tilde{f} ({\bf k}_q,{\bf k}_p)
\to {\cal F}^{-1} \{ \tilde{f}({\bf k}_q,{\bf k}_p)\}= 
f_C ({\bf q},{\bf p})$
that is defined by 
\be \label{uxt}
f_C ({\bf q},{\bf p}) =
{\cal F}^{-1} \{ \tilde{f} ({\bf k}_q,{\bf k}_p) \} 
= \frac{1}{(2\pi)^{6N}} 
\int^{+\infty}_{-\infty} dk_1 \ldots
\int^{+\infty}_{-\infty} dk_{6N} 
\ e^{i \sum^{6N}_{j=1}k_jx_j} 
\tilde{f} ({\bf k}_q,{\bf k}_p) .
\ee
}\\


For simplification, we will use notations
$f ({\bf n}_{\bf q},{\bf n}_{\bf p})$ for 
$f_L ({\bf n}_{\bf q},{\bf n}_{\bf p})$ and
$f({\bf q},{\bf p})$ for $f_C ({\bf q},{\bf p})$.
From the context it will be clear 
which function is considered.

We use the lattice-continuum transform operation 
${\cal T}_{L \to C}$ for 
the lattice functions $f ({\bf m}_q,{\bf m}_p,t)$,
for the product of two lattice functions
$( f ({\bf m}_q,{\bf m}_p,t) 
\, g ({\bf m}_q,{\bf m}_p,t) )$
and for lattice fractional derivatives
of the functions, i.e. $g ({\bf n}_q,{\bf n}_p,t) =
\mathbb{D}^{-}_{L} \left[ \alpha_j \atop j \right] 
f ({\bf m}_q,{\bf m}_p,t) $.
The operation ${\cal T}_{L \to C}$
can be applied not only for lattice functions 
but also for lattice operators.
The operation ${\cal T}_{L \to C}$
allows us to map of lattice derivative 
$\mathbb{D}_L^{-} \left[ \alpha_j \atop j \right]$
into a  phase-space  continuum fractional derivative 
$\mathbb{D}_C^{-} \left[ \alpha_j \atop j \right]$
that is defined in Appendix.  \\



{\bf Proposition}
{\it The lattice-continuum transform operation
${\cal T}_{L \to C}$ maps
the fractional Liouville equation 
for phase-space lattice (\ref{FLE-L}) 
with the lattice operators (\ref{Dalpha1}) 
into the fractional Liouville equation for 
phase-space continuum
\[
\frac{\partial \rho ({\bf q},{\bf p},t) }{\partial t} 
+ \sum^{3N}_{j=1} 
\mathbb{D}^{-}_{C,q} \left[ \alpha_j \atop j \right]
\, \Bigl( V_ j ({\bf q},{\bf p},t) 
\, \rho ({\bf q},{\bf p},t) \Bigr) +
\]
\be \label{FLE-C}
+ \sum^{6N}_{j=3N+1} 
\mathbb{D}^{-}_{C,p} \left[ \beta_j \atop j \right]
\, \Bigl( F_j ({\bf q},{\bf p},t) \,
\rho ({\bf q},{\bf p},t) \Bigr) =
J ({\bf q},{\bf p},t) ,
\ee
where $ \mathbb{D}^{-}_{C,q} \left[ \alpha_j \atop j \right]$
and $\mathbb{D}^{-}_{C,p} \left[ \beta_j \atop j \right]$
are the continuum fractional derivatives 
with respect to phase-space coordinate $q_j$ and $p_j$
that are defined by (\ref{CFD-1}) for $0<\alpha_j<1$, 
by equation (\ref{CFD-2}) for $\alpha_j>1$, and
by equation (\ref{CFD-3}) for integer odd $\alpha_j$ (see Appendix).
The functions $\rho ({\bf q},{\bf p},t)$,
$V_j ({\bf q},{\bf p},t)$, $F ({\bf q},{\bf p},t)$,
$J ({\bf q},{\bf p},t)$ are defined by the equations
\be \label{rhoT}
\rho ({\bf q},{\bf p},t) =
{\cal T}_{L \to C} \left( \rho ({\bf n}_q,{\bf n}_p,t) \right) ,
\quad 
J ({\bf q},{\bf p},t) =
{\cal T}_{L \to C} \left( J ({\bf n}_q,{\bf n}_p,t) \right) ,
\ee
\be 
V_j ({\bf q},{\bf p},t) =
{\cal T}_{L \to C} \left( V_j ({\bf n}_q,{\bf n}_p,t) \right) ,
\quad
F_j ({\bf q},{\bf p},t) =
{\cal T}_{L \to C} \left( F_j ({\bf n}_q,{\bf n}_p,t) \right) .
\ee
} \\


{\bf Proof.}

Applying the discrete Fourier transform 
${\cal F}_{\Delta}$ to the first term of 
the lattice Liouville equation (\ref{FLE-L}), we obtain
\be
{\cal F}_{\Delta} \left(
\frac{\partial \rho ({\bf n}_q,{\bf n}_p,t) }{\partial t} \right)
=
\frac{\partial {\cal F}_{\Delta} \left(\rho ({\bf n}_q,{\bf n}_p,t) \right) }{\partial t} =
\frac{\partial \hat{\rho} ({\bf k}_q,{\bf k}_p,t) }{\partial t} ,
\ee
where
\be
\hat{\rho} ({\bf k}_q,{\bf k}_p,t) =
{\cal F}_{\Delta} \left( 
\rho ({\bf n}_q,{\bf n}_p,t) \right) .
\ee
Analogously, we can see that
\be
{\cal T}_{L \to C} \left(
\frac{\partial \rho ({\bf n}_q,{\bf n}_p,t) }{\partial t} \right)
=
\frac{\partial {\cal T}_{L \to C} \left(\rho ({\bf n}_q,{\bf n}_p,t) \right) }{\partial t} =
\frac{\partial \rho ({\bf q},{\bf p},t) }{\partial t} ,
\ee
where $\rho ({\bf q},{\bf p},t)$ is defied by (\ref{rhoT}).


The discrete Fourier transform ${\cal F}_{\Delta}$ 
of the second term of (\ref{FLE-L}), 
gives
\[
{\cal F}_{\Delta} 
\left( \sum^{3N}_{j=1} \mathbb{D}_L^{-} \left[ \alpha_j \atop j \right] \ \Bigl( V_j ({\bf m}_q,{\bf m}_p,t)
\rho ({\bf m}_q,{\bf m}_p,t) \Bigr)
\right) =
\]
\[
= \sum^{3N}_{j=1} 
\left( \sum^{+\infty}_{n_j=-\infty} e^{-ik_j \, n_j \, a_j} \,
\mathbb{D}_L^{-} \left[ \alpha_j \atop j \right] 
\Bigl( V_j ({\bf m}_q,{\bf m}_p,t)
\rho ({\bf m}_q,{\bf m}_p,t) \Bigr) \right) =
\]
\[
= \sum^{3N}_{j=1} \frac{1}{a^{\alpha_j}_j} \left( 
\sum^{+\infty}_{n_j=-\infty} \
\sum^{+\infty}_{m_j=-\infty}
e^{-ik_j \,n_j \, a_j} \, K^{-}_{\alpha_j}(n_j-m_j) \, 
\Bigl( V_j ({\bf m}_q,{\bf m}_p,t)
\rho ({\bf m}_q,{\bf m}_p,t) \Bigr) \right) =
\]
\[
= \sum^{3N}_{j=1} \frac{1}{a^{\alpha_j}_j} \left( 
\sum^{+\infty}_{n^{\prime}_j=-\infty} 
e^{-ik_j \, n^{\prime}_j \, a_j} K^{-}_{\alpha_j}(n^{\prime}_j) \
\sum^{+\infty}_{m_j=-\infty } 
\Bigl( V_k ({\bf m}_q,{\bf m}_p,t)
\rho ({\bf m}_q,{\bf m}_p,t) \Bigr) 
e^{-ik_j \, m_j \, a_j}\right) = 
\]
\be \label{C9}
= \sum^{3N}_{j=1} \frac{1}{a^{\alpha_j}_j} \left( 
\hat{K}^{-}_{\alpha_j} (k_j \, a_j) \ 
\Bigl( \hat{V}_j ({\bf k}_q,{\bf k}_p,t) *
\hat{\rho} ({\bf k}_q,{\bf k}_p,t) \Bigr)_q \right),
\ee
where $\hat{K}^{-}_{\alpha_j} (k_j \, a_j)$ 
is defined by (\ref{Kn1-}), and $n^{\prime}_j = n_j - m_j$.
The symbol $( \ * \ )_q$ denotes 
the convolution with respect to ${\bf k}_q$ that
is defined by the equation
\be
\Bigl( \hat{f} ({\bf k}_q,{\bf k}_p,t) *
\hat{g} ({\bf k}_q,{\bf k}_p,t) \Bigr)_q =
\int_{-k_{0}/2}^{+k_{0}/2} \prod^{3N}_{j=1} dk^{\prime}_j \,
\hat{f} ( {\bf k}^{\prime}_q ,{\bf k}_p,t) *
\hat{g} ({\bf k}_q -{\bf k}^{\prime}_q,{\bf k}_p,t) .
\ee

Similarly, the transform ${\cal F}_{\Delta}$ of 
the third term of (\ref{FLE-L}) gives
\be \label{C3a2}
{\cal F}_{\Delta} \left( \sum^{6N}_{j=3N+1} 
\mathbb{D}_L^{-} \left[ \alpha_j \atop j \right] 
\rho ({\bf m}_q,{\bf m}_p,t)
\right) = 
\sum^{6N}_{j=3N+1} \frac{1}{a^{\beta_j}_j} \left( 
\hat{K}^{-}_{\beta_j} (k_j \, a_j) \ 
\Bigl( \hat{V}_j ({\bf k}_q,{\bf k}_p,t) *
 \hat{\rho} ({\bf k}_q,{\bf k}_p,t) \Bigr)_p \right),
\ee
where $( \ * \ )_p$ denotes 
the convolution with respect to ${\bf k}_p$
that is defined by the equation
\be
\Bigl( \hat{f} ({\bf k}_q,{\bf k}_p,t) *
\hat{g} ({\bf k}_q,{\bf k}_p,t) \Bigr)_p =
\int_{-k_{0}/2}^{+k_{0}/2} \prod^{6N}_{j=3N+1} dk^{\prime}_j \,
\hat{f} ( {\bf k}_q ,{\bf k}^{\prime}_p,t) *
\hat{g} ({\bf k}_q,{\bf k}_p-{\bf k}^{\prime}_p,t) .
\ee

As a result, the Liouville equation has the form
\[
\frac{\partial \hat{\rho} ({\bf k}_q,{\bf k}_p,t) }{\partial t} 
+ \sum^{3N}_{j=1} \hat{K}^{-}_{\alpha_j} (k_j \, a_j) \,
\Bigl( \hat{V}_j ({\bf k}_q,{\bf k}_p,t) *
 \hat{\rho} ({\bf k}_q,{\bf k}_p,t) \Bigr)_q +
\]
\be \label{FLE-2}
+ \sum^{6N}_{j=3N+1} \hat{K}^{-}_{\beta_j} (k_j \, a_j) \, 
\Bigl( \hat{F}_j ({\bf k}_q,{\bf k}_p,t) *
\hat{\rho} ({\bf k}_q,{\bf k}_p,t) \Bigr)_p =
\hat{J} ({\bf k}_q,{\bf k}_p,t) ,
\ee
where the symbols $( \ * \ )_q$ and $( \ * \ )_p$
denote the convolution with respect to ${\bf k}_q$
and ${\bf k}_p$ respectively.


Then we use 
\be
\hat{K}^{-}_{\alpha_j}(a_j \, k_j) = 
i \, \operatorname{sgn}(k_j) \, |a_j\, k_j|^{\alpha_j} ,
\quad (j =1, \ldots , 3N) , 
\ee
\be
\hat{K}^{-}_{\beta_j}(a_j \, k_j) = 
i \, \operatorname{sgn}(k_j) \, |a_j\, k_j|^{\beta_j} ,
\quad (j =3N+1, \ldots , 6N) . 
\ee
The limit $a_j \rightarrow 0$ gives
\be 
\tilde{K}^{-}_{\alpha_j}(k_j) = \lim_{a_j \to 0} \, 
\frac{1}{a^{\alpha_j}_j} \,
\hat{K}^{-}_{\alpha_j} (k_j \, a_j) =
 i \, k_j \, |k_j|^{\alpha_j -1} \quad (j =1, \ldots , 3N) ,
\ee
\be 
\tilde{K}^{-}_{\beta_j}(k_j) = \lim_{a_j \to 0} \, 
\frac{1}{a^{\beta_j}_j} \,
\hat{K}^{-}_{\beta_j} (k_j \, a_j) =
i \, k_j \, |k_j|^{\beta_j -1} \quad (j =3N+1, \ldots , 6N) .
\ee


The "hat"-kernel $\hat{K}^{-}_{\alpha_j} (k_j)$ 
is the discrete Fourier transform 
${\cal F}_{\Delta}$ of the kernel of lattice operator. 
The equation that defines $\hat{K}^{-}_{\alpha_j} (k_j)$ 
has the form
\be 
{\cal F}_{\Delta} \left( \mathbb{D}_L^{-} 
\left[ \alpha_j \atop j \right] \,
f ({\bf m}_q,{\bf m}_p,t) \right) =
\frac{1}{a^{\alpha_j}_j} \hat{K}^{-}_{\alpha_j} 
(k_j \, a_j) \, \hat{f} ({\bf k}_q,{\bf k}_p,t) ,
\ee 
where 
$\hat{f} ({\bf k}_q,{\bf k}_p,t) = 
{\cal F}_{\Delta} \{ f ({\bf m}_q,{\bf m}_p,t) \}$.

The "tilde"-kernel $\tilde{K}^{-}_{\alpha_j} (k_j)$ 
is the Fourier integral transform ${\cal F}$
of the continuum derivative.
The equation that defines 
$\tilde{K}^{-}_{\alpha_j} (k_J)$ is 
\be 
{\cal F} \left( \mathbb{D}_C^{-} 
\left[ \alpha_j \atop j \right] \,
f ({\bf q},{\bf p},t) \right) =
\frac{1}{a^{\alpha_j}_j} \tilde{K}^{-}_{\alpha_j} 
(k_j \, a_j) \, \tilde{f} ({\bf k}_q,{\bf k}_p,t) ,
\ee 
where $\tilde{f} ({\bf k}_q,{\bf k}_p,t) 
= {\cal F} \{ f ({\bf q},{\bf p},t) \}$. 


As a result, the limit $a_j \rightarrow 0$ for (\ref{FLE-2}) 
gives the Liouville equation in the form
\[
\frac{\partial \tilde{\rho} ({\bf k}_q,{\bf k}_p,t) }{\partial t} 
+ \sum^{3N}_{j=1} \tilde{K}^{-}_{\alpha_j} (k_j) \,
\Bigl( \tilde{V}_j ({\bf k}_q,{\bf k}_p,t) *
 \tilde{\rho} ({\bf k}_q,{\bf k}_p,t) \Bigr)_q +
\]
\be \label{FLE-3}
+ \sum^{6N}_{j=3N+1} \tilde{K}^{-}_{\beta_j} (k_j) \, 
\Bigl( \tilde{F}_j ({\bf k}_q,{\bf k}_p,t) *
\tilde{\rho} ({\bf k}_q,{\bf k}_p,t) \Bigr)_p =
\tilde{J} ({\bf k}_q,{\bf k}_p,t) ,
\ee
where 
\be 
\tilde{K}^{-}_{\alpha_j}(k_j) = i \, k_j \, |k_j|^{\alpha_j -1} , 
\quad (j=1, \ldots , 3N) ,
\ee
\be
\tilde{K}^{-}_{\beta_j}(k_j) = i \, k_j \, |k_j|^{\beta_j -1} , 
\quad (j=3N+1, \ldots , 6N) ,
\ee
and
\be
\tilde{\rho} ({\bf k}_q,{\bf k}_p,t) = \operatorname{Lim} 
\hat{\rho} ({\bf k}_q,{\bf k}_p,t) 
\quad
\tilde{J} ({\bf k}_q,{\bf k}_p,t) = \operatorname{Lim} 
\hat{J} ({\bf k}_q,{\bf k}_p,t) . 
\ee

The inverse Fourier transform of (\ref{FLE-3}) 
gives the fractional Liouville equation (\ref{FLE-C}).
As a result, we prove that 
the lattice fractional Liouville equation (\ref{FLE-L})
is transformed by the operation ${\cal T}_{L \to C}$ 
into the continuum fractional Liouville equation (\ref{FLE-C}).

This ends the proof. \\


\section{Fractional Liouville equation for Hamiltonian systems}

For wide class of physical $N$-particle systems we can use 
\be \label{Vj-0}
V_j ({\bf q},{\bf p},t) = V_j ({\bf p}) , 
\quad (j=1, \ldots , 3N) 
\ee
\be \label{Fj-0}
F_j ({\bf q},{\bf p},t) = F_j ({\bf q},t) , 
\quad (j=3N+1, \ldots , 6N) ,
\ee
For Hamiltonian systems with potential forces, 
we have
\be \label{Vj-1}
V_j ({\bf p}) = \frac{\partial T({\bf p})}{\partial p_j} , 
\quad (j=1, \ldots , 3N) 
\ee
\be \label{Fj-1}
F_j ({\bf q}) = - \frac{\partial U({\bf q})}{\partial q_j} ,
\quad (j=3N+1, \ldots , 6N) ,
\ee
where the Hamiltonian is 
$ H({\bf q},{\bf p}) = T({\bf p})+U({\bf q})$, 
the function $U({\bf q})$ is 
the generalized potential of the force,
and $ T({\bf p})$ is the generalized kinetic energy term.
For simple case, we have $V_j ({\bf p})= p_j/M$.
Conditions (\ref{Vj-0}) and (\ref{Fj-0}) mean that $V_j$ are
components of the particle velocity,
and $F_j$ are components of the force vector
that is independent of the particle momenta. 
In the case (\ref{Vj-0}) and (\ref{Fj-0}), 
the fractional Liouville equation for 
phase-space continuum can be written in the form
\[
\frac{\partial \rho ({\bf q},{\bf p},t) }{\partial t} 
+ \sum^{3N}_{j=1} \, V_ j ({\bf p}) \,
\mathbb{D}^{-}_{C,q} \left[ \alpha_j \atop j \right]
\, \rho ({\bf q},{\bf p},t) +
\]
\be \label{FLE-C-2}
+ \sum^{6N}_{j=3N+1} \, F_j ({\bf q},t) \,
\mathbb{D}^{-}_{C,p} \left[ \beta_j \atop j \right]
\, \rho ({\bf q},{\bf p},t) =
J ({\bf q},{\bf p},t) .
\ee


As a special case, we also can consider 
fractional Hamiltonian systems 
\cite{FHS-1,FHS-2,TarasovSpringer}, where
\be
V_j ({\bf q},{\bf p},t) =
\mathbb{D}^{-}_{C,p} \left[ \beta_{j+3N} \atop j+3N \right] \,
H_{\alpha,\beta} ({\bf q},{\bf p}) ,
\quad (j=1, \ldots , 3N) ,
\ee
\be
F_j ({\bf q},{\bf p},t) = -
\mathbb{D}^{-}_{C,q} \left[ \alpha_{j-3N} \atop j - 3N \right] \,
H_{\alpha,\beta} ({\bf q},{\bf p}) , 
\quad (j=3N+1, \ldots , 6N) ,
\ee
where $ H_{\alpha,\beta} ({\bf q},{\bf p}) $
is the generalized Hamiltonian function \cite{FHS-1}.
In this case, the fractional Liouville equation 
can be represented in the form
\be \label{FLE-C-3}
\frac{\partial \rho ({\bf q},{\bf p},t) }{\partial t} +
\{H_{\alpha,\beta} ({\bf q},{\bf p}), \rho ({\bf q},{\bf p},t)\}_{\alpha,\beta}
= J ({\bf q},{\bf p},t) ,
\ee
where $ \{ \ , \ \}_{\alpha,\beta}$ is 
the fractional Poisson brackets
\[
\{f ({\bf q},{\bf p}), g ({\bf q},{\bf p})\}_{\alpha,\beta} =
\sum^{3N}_{j=1} \Bigl(
\mathbb{D}^{-}_{C,p} \left[ \beta_{3N+j} \atop 3N+j \right] 
\, f ({\bf q},{\bf p},t) \,
\mathbb{D}^{-}_{C,q} \left[ \alpha_j \atop j \right] 
\, g ({\bf q},{\bf p},t)
-
\]
\be \label{PB}
- \mathbb{D}^{-}_{C,q} \left[ \alpha_j \atop j \right] 
\, f ({\bf q},{\bf p},t) \,
\mathbb{D}^{-}_{C,p} \left[ \beta_{3N+j} \atop 3N+j \right] 
\, g ({\bf q},{\bf p},t)
\Bigr) .
\ee
If all $\alpha_j=1$ and $\beta_j=1$, then
(\ref{PB}) is the usual Poisson brackets,
and (\ref{FLE-C-3}) is the usual Liouville equation
for classical Hamiltonian systems.

If we use $({\bf n}_{\bf q},{\bf n}_{\bf p})$ instead of 
$({\bf q},{\bf p})$, 
and $\mathbb{D}^{-}_{L,q}$, $\mathbb{D}^{-}_{L,p}$
instead of  
$\mathbb{D}^{-}_{C,q}$, $\mathbb{D}^{-}_{C,p}$
in equations (\ref{Vj-0}-\ref{PB}), 
then we get the correspondent lattice analogs of
the considered equations.


\section{Fractional Liouville equation for nonlocal media}

As an example of application of the fractional Liouville
equation with conjugate Riesz fractional derivatives,
we consider description of non-local plasma-like continuum.

For simplification, we consider 
an isotropic collisionless nonlocal plasma-like media, 
where all $\alpha_j =\alpha$ and $\beta_j=1$,
$J ({\bf q},{\bf p},t)=0$ and $V_j ({\bf p}) = p_j/M$.
In this case, the $N$-particle distribution function 
$\rho ({\bf q},{\bf p},t) $ is the product of 
one-particle reduced 
distribution function $\rho_1 ({\bf x},{\bf p},t)$, 
\be
\rho ({\bf q},{\bf p},t) = \prod^{N}_{s=1} 
\rho_1 ({\bf x}_s,{\bf p}_s,t) ,
\ee
where we use usual space coordinates 
${\bf x}_s=\sum^3_{j=1} {\bf e}_j x_{sj}$
and ${\bf p}_s=\sum^3_{j=1} {\bf e}_j p_{sj}$
instead of the generalized phase-space coordinates 
${\bf q} = (q_1, \ldots ,q_{3N})$ and
${\bf p} = (p_1, \ldots ,p_{3N})$.
Here ${\bf e}_j$ are the basis vectors
of the Cartesian coordinate system in $\mathbb{R}^3$.
The distribution function $\rho_1 ({\bf x},{\bf p},t)$ 
describes a probability density to find the particle 
in the phase volume $d^{3}{\bf x} d^{3}{\bf p}$.

In this case, the fractional Liouville equation (\ref{FLE-C-2}) 
for the one-particle distribution function $\rho_1$
takes the form 
\be \label{Liouv3}
\frac{\partial \rho }{\partial t} 
+ \sum^{3}_{j=1} \, V_ j ({\bf p}) \,
\mathbb{D}^{-}_{C,x} \left[ \alpha \atop j \right]
\, \rho_1 +
\sum^{3}_{j=1} \, F_{3+j} ({\bf x},t) \,
\frac{\partial \rho_1}{\partial p_j} = 0 ,
\ee
where we use
\be
\mathbb{D}^{-}_{C,p} \left[ 1 \atop 3N+j \right] =
\frac{\partial}{\partial p_j} .
\ee
Let us apply this Liouville equation 
with space-fractional derivatives 
$\mathbb{D}^{-}_{C,x} \left[ \alpha \atop j \right]$
to describe properties of nonlocal media.

In the absence of the force field, 
the Liouville equation (\ref{Liouv3}) gives
\be \label{Liouv4}
\frac{\partial \rho }{\partial t} 
+ \sum^{3}_{j=1} \, V_ j ({\bf p}) \,
\mathbb{D}^{-}_{C,x} \left[ \alpha \atop j \right]
\, \rho = 0 .
\ee
The solution of this equation will be denoted by 
$\rho_0({\bf x},{\bf p},t)$.
For a weak force field, we can use 
the charge distribution function in the form 
\be
\rho_1=\rho_0 + \delta \rho ,
\ee
where $\rho_0$ is the stationary isotropic homogeneous 
distribution function unperturbed by the fields, and
$\delta \rho$ describes the change of $\rho_0$ by the fields. 
In the linear approximation with respect 
to field perturbation, we have
\be \label{Liouv5}
\frac{\partial \delta \rho }{\partial t} 
+ \sum^{3}_{j=1} \left( V_ j ({\bf p}) \,
\mathbb{D}^{-}_{C,x} \left[ \alpha \atop j \right]
\, \delta \rho +
F_{3+j} ({\bf x},t) \,
\frac{\partial \rho_0}{\partial p_j} \right) = 0 .
\ee
If we consider plasma-like media, then 
the force is the Lorentz force
${\bf F}= q {\bf E} + q [{\bf V},{\bf B}]$,
where $q$ is the charge of a particle that moves with 
velocity ${\bf V}= {\bf p}/M$ 
in the presence of an electric field 
${\bf E}$ and a magnetic field ${\bf B}$.

In an isotropic media, the distribution function 
depends only on the momentum, i.e. $\rho_0=\rho_0(|{\bf p}|)$. 
In this case, the direction of the vector 
${\bf e}_j \mathbb{D}^{-}_{C,x} \left[ \alpha \atop j \right]\rho_0$ coincides with the vector ${\bf p} = M {\bf V}$, 
and its scalar product with $[{\bf V},{\bf B}]$ 
is equal to zero. 
Therefore, the magnetic field does not affect 
the distribution function
in the linear approximation (\ref{Liouv5}). 
As a result, we have
\be \label{Liouv6}
\frac{\partial \delta \rho }{\partial t} 
+ \sum^{3}_{j=1} \left( \, V_ j ({\bf p}) \,
\mathbb{D}^{-}_{C,x} \left[ \alpha \atop j \right]
\, \delta \rho +
q \, E_{j} ({\bf x},t)\,
\frac{\partial \rho_0}{\partial p_j} \right) = 0 .
\ee
The Fourier transform with respect to space and time gives 
\be \label{Liouv7}
i \, \Bigl( \sum^{3}_{j=1} \operatorname{sgn}(k_j) \, 
|k_j|^{\alpha} V_j 
- \omega \Bigr) \delta \rho + 
q \sum^{3}_{j=1} 
\left( E_j \, \frac{\partial \rho_0}{\partial p_j} \right) =0 .
\ee

If we take the $X$-axis along ${\bf k}$, then we have
${\bf k}=(k_x,0,0)$, and
\[ \sum^{3}_{j=1} \operatorname{sgn}(k_j) \, |k_j|^{\alpha} V_j 
= |k_x|^{\alpha} V_x , \]
where $k_x> 0$ and $\operatorname{sgn}(k_x)=1$, 
since we use ${\bf e}_x={\bf k}/|{\bf k}|$,
and $k_x=|{\bf k}|$.
In this case, equation (\ref{Liouv7}) gives
\be \label{Liouv8}
i \, \Bigl( |k_x|^{\alpha} V_x 
- \omega \Bigr) \delta \rho + 
q \sum^{3}_{j=1} 
\left( E_j \, \frac{\partial \rho_0}{\partial p_j} \right) =0 .
\ee
Then we have
\be \label{delta-rho}
\delta \rho = - \frac{q}{i \, 
\Bigl( |k_x|^{\alpha} V_x - \omega \Bigr)} \, \sum^{3}_{j=1} 
\left( E_j \, \frac{\partial \rho_0}{\partial p_j} \right) . 
\ee

In an unperturbed plasma-like media, 
the charge density is equal zero. 
The charge density that is perturbed by the field is
\be \label{rho-ch}
\rho_{charge} = 
q \iiint^{+\infty}_{-\infty} \delta \rho \, d^3 {\bf p} =
i q^2 \iiint^{+\infty}_{-\infty} 
\frac{1}{ |k_x|^{\alpha} V_x - \omega } \, 
\sum^{3}_{j=1} 
\left( E_j \, \frac{\partial \rho_0}{\partial p_j} \right) 
\, d^3 {\bf p} ,
\ee
where $\rho_{charge}$ is the bound charge density.
The electric polarization vector ${\bf P}$ 
is defined by the relations 
\be
\operatorname{div} {\bf P} = - \rho_{charge} ,
\ee
which has the Fourier transform in the form
\be \label{Prho}
i ({\bf k},{\bf P})= - \rho_{charge} .
\ee
The polarization ${\bf P}$ defines 
the electric displacement field ${\bf D}$ 
by the equation 
${\bf D} = \varepsilon_0 {\bf E} + {\bf P}$,
where $\varepsilon_0$ is the electric permittivity.
Let the field ${\bf E}$ be parallel to ${\bf k}$. 
Then ${\bf P}$ be parallel to ${\bf k}$, and 
\be \label{bfP}
{\bf P}= \Bigl( \varepsilon_{\parallel} (|{\bf k}|) - \varepsilon_0 \Bigr) \, {\bf E} ,
\ee
where $\varepsilon_{\parallel} (|{\bf k}|)$ is the longitudinal permittivity.
Substitution of (\ref{rho-ch}) and (\ref{bfP}) into (\ref{Prho}) gives 
\be
(\varepsilon_{\parallel} (|{\bf k}|) - \varepsilon_0) \, 
({\bf k} , {\bf E}) =
- q^2 \iiint^{+\infty}_{-\infty} 
\frac{1}{ |k_x|^{\alpha} V_x - \omega } \, \sum^{3}_{j=1} 
\left( E_j \, \frac{\partial \rho_0}{\partial p_j} \right) 
\, d^3 {\bf p} ,
\ee
where $({\bf k} , {\bf E})$ is the scalar product of vectors
${\bf k}$ and ${\bf E}$.
Since we take the $X$-axis along the vector ${\bf k}$, 
then ${\bf E}=(E_x,0,0)$, and
\be
({\bf k} , {\bf E})= k_x E_x= |{\bf k}| E_x , \quad
\sum^{3}_{j=1} 
\left( E_j \, \frac{\partial \rho_0}{\partial p_j} \right)=
E_x \, \frac{\partial \rho_0}{\partial p_x} ,
\ee
where we can use $|{\bf k}|$ instead of $k_x$.

Using the distribution function
\be
\rho_0(p_x) = \iint^{+\infty}_{-\infty} 
\rho_0 (|{\bf p}|) \, dp_y \, dp_z ,
\ee
we get the following equation to calculate
the longitudinal permittivity 
\be \label{e1}
\varepsilon_{\parallel} (|{\bf k}|) = 
\varepsilon_0 - \frac{q^2}{ |{\bf k}| }
\int^{+\infty}_{-\infty} 
\frac{1}{ |k_x|^{\alpha} p_x/M - \omega } \, 
\frac{\partial \rho_0(p_x)}{\partial p_x} \, d p_x ,
\ee
where $|k_x|=|{\bf k}|$ can be used. 
For isotropic homogeneous case, 
we can use an equilibrium distribution $\rho_0(p_x)$.


Let us consider a plasma-like medium with 
the equilibrium Maxwell's distribution
\be
\rho_0 (p_x) = \frac{N_q}{\sqrt{2 \pi M k_B T}} 
\exp \left( - \frac{p^2_x}{2 M k_B T}\right) ,
\ee
where $k_B$ is the Boltzmann constant. Then
\be
\frac{\partial \rho_0(p_x)}{\partial p_x} = 
-\frac{2 p_x N_q}{\sqrt{\pi} \, (2 M k_B T)^{3/2}} \exp \left( - \frac{p^2_x}{2 M k_B T}\right) ,
\ee
where $N_q$ is the total number of particles per unit volume.

Using $k_x=|{\bf k}|$, equation (\ref{e1}) 
can be rewritten in the form
\be \label{71}
\varepsilon_{\parallel} (|{\bf k}|) =\varepsilon_0 + \frac{q^2 N_q}{ |{\bf k}|^{1+\alpha} }
\frac{2 M}{\sqrt{\pi} (2 M k_B T)^{3/2}}
\int^{+\infty}_{-\infty} \frac{p_x}{ p_x - M \omega / |{\bf k}|^{\alpha} - i 0} \, 
\exp \left( - \frac{p^2_x}{2m k_B T}\right) \, d p_x .
\ee
Using new variables 
\be \label{zxi}
z= \frac{p_x}{\sqrt{2M k_B T}} , \quad 
\xi = \sqrt{\frac{M}{2 k_B T} } \cdot \frac{\omega}{|{\bf k}|^{\alpha}} ,
\ee
equation (\ref{71}) can be represented in the form
\be \label{eq-int}
\varepsilon_{\parallel} (|{\bf k}|) = \varepsilon_0 + \frac{q^2}{ |{\bf k}|^{1+\alpha} }
\frac{1}{ \sqrt{\pi} k_B T }
\int^{+\infty}_{-\infty} \frac{z \, e^{-z^2}}{ z - \xi - i 0} \, d z ,
\ee
where 
\be \label{int}
\int^{+\infty}_{-\infty} \frac{z \, e^{-z^2}}{ z - \xi - i 0} \, d z = \sqrt{\pi}+
P.V. \int^{+\infty}_{-\infty} \frac{\xi e^{-z^2}}{ z - \xi} \, d z + i \pi \xi e^{-\xi^2} .
\ee
It should be noted that ${\bf k}$, ${\bf x}$, $k_x$ and $x_j$ 
are dimensionless variables.


We consider equation (\ref{eq-int}) for two cases
that are characterized by the large and small values of 
the variable $\xi$. 


1) For small values $\xi \ll 1$, we have
\[ P.V. \int^{+\infty}_{-\infty} 
\frac{x e^{-z^2}}{ z - \xi} \, d z = 
P.V. \int^{+\infty}_{-\infty} 
\frac{\xi e^{-(y+\xi)^2}}{y} \, d y 
=
\]
\[
P.V. \int^{+\infty}_{-\infty} e^{-y^2} \Bigl( y^{-1} \xi -
2 \xi^2 - ( y^{-1} + 2y) \xi^3 + (2 - (4/3) y^2) \xi^4 +
\ldots \Bigr) \, d y = \]
\be \label{Eq-app-1}
= -2 \sqrt{\pi} \, \xi^2 + \sqrt{\pi} \, \xi^4 + \ldots \ \quad
(\xi \ll 1) ,
\ee
where $y=z-\xi$, and 
we take into account that the integrals of 
the odd terms in $y$ are zero.
Substitution of (\ref{Eq-app-1}) 
and (\ref{int}) into (\ref{eq-int}) gives
\be \label{Final-12}
\varepsilon_{\parallel} (|{\bf k}|) = 
\varepsilon_0 + \frac{q^2 N_q}{k_B T \, |{\bf k}|^{1+\alpha} }
- \frac{q^2 N_q M \omega^2}{ k^2_B T^2 \, |{\bf k}|^{3\alpha+1} } +
\frac{q^2 N_q M^2 \omega^4}{ 4 k^3_B T^3 \, |{\bf k}|^{5\alpha+1} } + \ldots \ .
\ee
In the case $\xi \ll 1$, the imaginary part of 
the permittivity $\varepsilon_{\parallel} (|{\bf k}|)$ 
is relatively small but it is not exponentially small, 
because of the smallness of the phase volume, where 
the condition $|{\bf k}|^{\alpha} p_x/M - \omega =0$ holds.


2) For large values $\xi \gg 1$, we have
\[ 
P.V. \int^{+\infty}_{-\infty} \frac{x e^{-z^2}}{ z - \xi} \, d z 
=
- \int^{+\infty}_{-\infty} \frac{e^{-z^2}}{ 1- z/\xi } \, d z = 
- \int^{+\infty}_{-\infty} e^{-z^2} \left(1+\sum^{\infty}_{m=1} 
\left(\frac{z}{\xi} \right)^m \right) \, d z =
\]
\be \label{Eq-app-2}
= - \sqrt{\pi} - \frac{\sqrt{\pi}}{2 \xi^2}- \frac{3\sqrt{\pi}}{4 \xi^4}- \ldots
\quad (\xi \gg 1) ,
\ee
where we take into account that 
the integrals of the odd terms in $z$ are zero also. 
Substituting (\ref{Eq-app-2}) and (\ref{int}) 
into (\ref{eq-int}), we get
\be \label{Final-22}
\varepsilon_{\parallel} (|{\bf k}|) = \varepsilon_0 -
\frac{q^2 N_q}{ M \omega^2} \, |{\bf k}|^{\alpha-1} -
\frac{3 q^2 N_q k_B T}{M^2 \omega^4} \, |{\bf k}|^{3 \alpha-1} + \ldots \ .
\ee
For Maxwell's distribution, 
an exponentially small part of the charged particles has 
the velocity $V_x = \omega/ |{\bf k}| \gg V_T= \sqrt{k_B T/m}$,
where $V_T$ is the average velocity of charged particles. 
Therefore the imaginary part of the permittivity
$\varepsilon_{\parallel} (|{\bf k}|)$ is exponentially small. 

Equations (\ref{Final-12}) and (\ref{Final-22})
can be used to obtain the scalar potentials of electric field 
for the fractional nonlocal 
plasma-like media \cite{TFLE-1,AP2013}.


\section{Conclusion}

In this paper, the Liouville equation 
for unbounded homogeneous phase-space 
lattice with long-range jumps is suggested.
Using the methods proposed in \cite{JPA2006,JMP2006,JPA2014}, 
we prove that the continuous limit transforms
the suggested Liouville equation for lattice phase-space
into the fractional nonlocal Liouville equation 
for continuum phase-space.
This fractional Liouville equation contains
the generalized conjugate Riesz derivatives 
on non-integer orders with respect to phase-space coordinates.
As an example, we consider an application
of the fractional Liouville equation with 
the Riesz derivatives of non-integer orders
to describe properties of plasma-like nonlocal media.

We assume that the lattice fractional Fokker-Planck 
equation, which is suggested in \cite{JSM2014},
can be obtained from the phase-space lattice Liouville equation 
that is proposed in this paper.
The lattice Bogoliubov hierarchy equations 
can be easily derived from the suggested 
lattice Liouville equation
by using the reduced distribution functions.
The correspondent lattice hydrodynamic approximation 
can also be obtained from 
the lattice Bogoliubov hierarchy equations.
It allows us to formulate fractional-nonlocal
statistical mechanics \cite{TFLE-5,TarasovSpringer} 
on the lattice phase-space.
The suggested lattice-continuum transform operator
allows us to have a correspondence
between lattice fractional statistical mechanics
and fractional nonlocal statistical theory.

The fractional Bogoliubov hierarchy equations 
with derivatives of non-integer orders with respect to
phase-space coordinates can be obtained 
from the suggested fractional Liouville equation 
with Riesz fractional derivatives
in the same way as was done for the Liouville equation with Caputo derivatives in \cite{TFLE-5,TFLE-6}.
The Vlasov equation and the Fokker-Planck equation
with the fractional derivatives of the Riesz type
can be also obtained from fractional Liouville equations
\cite{JSM2014}.
All these equations form a basis of fractional
nonlocal statistical mechanics.



\newpage

\section*{Appendix: Continuum fractional derivative 
of the Riesz type}

Let us consider the continuum fractional derivative 
$\mathbb{D}_C^{-} \left[ \alpha_j\atop j \right]$
of the Riesz type that has the property
\be \label{FdelZ-Def2}
{\cal F} \left( 
\mathbb{D}_C^{-} \left[ \alpha_j \atop j \right] \, 
f ({\bf x}) \right) ({\bf k})
 = i \, \operatorname{sgn}(k_j) \, 
|k_j|^{\alpha_j} ({\cal F} f)({\bf k}) 
\quad (\alpha_j >0). 
\ee
For $0<\alpha_j<1$ the operator 
$\mathbb{D}_C^{-} \left[ \alpha_j \atop j \right]$ 
can be considered as the conjugate Riesz derivative 
\cite{Uch} with respect to $x_j$.
Therefore, the operator (\ref{FdelZ-Def2})
can be called a generalized conjugate 
derivative of the Riesz type.


For $0<\alpha_j<1$ the fractional operator (\ref{FdelZ-Def2})
can be defined by the equation
\be \label{CFD-1}
\mathbb{D}_C^{-} \left[ \alpha_j \atop j \right] 
f ({\bf x}) = \frac{\partial}{\partial x_j} \, 
\int_{\mathbb{R}^1} R_{1-\alpha_j}(x_j-z_j) \,
f ({\bf x}+ (z_j-x_j) \, {\bf e}_j) \, dz_j, 
\quad (0<\alpha_j <1) ,
\ee
where ${\bf e}_j$ is the basis of the Cartesian coordinate system.
The function $R_{\alpha_j}(x)$ is the Riesz kernel, that is defined by 
\be 
R_{\alpha_j}(x) = 
\left\{
\begin{array}{cc}
\gamma^{-1}_1(\alpha_j) |x|^{\alpha_j-1} 
& \alpha_j\ne 2n+1, \quad n \in \mathbb{N} , \\
& \\
- \gamma^{-1}_1(\alpha_j) |x|^{\alpha_j-1} \ln |x| 
& \alpha_j=2 n+1 , \quad n \in \mathbb{N} .
\end{array}
\right.
\ee
The constant $\gamma_1(\alpha_j)$ has the form
\be 
\gamma_1(\alpha_j)=
\left\{
\begin{array}{cc}
2^{\alpha_j} \pi^{1/2}\Gamma(\alpha/2)/ \Gamma((1-\alpha_j)/2) &
\alpha_j\ne 2n+1, \\
& \\
(-1)^{(1-\alpha_j)/2}2^{\alpha_j-1} \pi^{1/2} \;
\Gamma(\alpha/2) \; 
\Gamma( 1+[\alpha_j-1]/2) 
& \alpha_j=2n+1 ,
\end{array}
\right.
\ee
where $N \in \mathbb{N}$ and $\alpha_j\in \mathbb{R}_{+}$.
Note the distinction between 
the continuum fractional derivatives 
$\mathbb{D}_C^{-} \left[ \alpha_j \atop j \right]$
and the Riesz potential consists in the use of 
$|k_j|^{-\alpha_j}$ instead of $|{\bf k}|^{-\alpha_j}$.


For $\alpha_j>1$ the fractional operator (\ref{FdelZ-Def2})
can be defined by the equation
\be \label{CFD-2}
\mathbb{D}_C^{-} \left[ \alpha_j \atop j \right] \, 
f({\bf x})
= \frac{1}{d_1(m,\alpha_j-1)} \frac{\partial}{\partial x_j}
\int_{\mathbb{R}^1} \, 
\frac{1}{|z_j|^{\alpha_j}} \,
(\Delta^m_{z_j} f)({\bf x}) \, dz_j , 
\quad (1 < \alpha_j < m +1) ,
\ee
where $(\Delta^m_{z_j} u)({\bf x})$ is a finite difference of
order $m$ of a function $f ({\bf x})$ with the vector step 
${\bf z}_j= z_j \, {\bf e}_j \in \mathbb{R}^N$
for the point ${\bf x}=\sum^{N}_{j=1} x_j \, {\bf e}_j 
\in \mathbb{R}^N$.
The centered difference
\be \label{cent}
(\Delta^m_{z_j} f)({\bf x}_j) =\sum^m_{n=0} (-1)^n 
\frac{m!}{n! \, (m-n)!} \, 
f ({\bf x}-(m/2-n) \, z_j \, {\bf e}_j) . 
\ee
The constant $d_1(m,\alpha_j)$ is defined by
\[ d_1(m,\alpha_j) =\frac{\pi^{3/2} A_m(\alpha_j)}{2^{\alpha_j} 
\Gamma(1+\alpha/2) \Gamma((1+\alpha_j)/2) \sin (\pi \alpha/2)} , \]
where 
\[ A_m(\alpha_j)=2\, \sum^{[m/2]}_{s=0} (-1)^{s-1} 
\frac{m!}{s!(m-s)!} \, (m/2- s)^{\alpha_j} \]
for the centered difference (\ref{cent}).
The constant $d_1(m,\alpha_j)$ is different from zero
for all $\alpha_j>0$ in the case of an even $m$ and 
centered difference $(\Delta^m_{i} u)$ (see Theorem 26.1 in \cite{SKM}).
Note that the derivative (\ref{CFD-2})
does not depend on the choice of $m>\alpha_j-1$.
Therefore, we can always choose an even number $m$ 
so that it is greater than $\alpha_1-1$,
and then we can use the centered difference (\ref{cent}) 
for all positive real values of $\alpha_j$.


For integer odd values of $\alpha_j$, we have
\be \label{CFD-3}
\mathbb{D}_C^{-} \left[ 2m+1 \atop j \right] \, f ({\bf x}) = 
(-1)^m \, 
\frac{\partial^{2m+1} f ({\bf x})}{\partial x_j^{2m+1} } , 
\quad ( m \in \mathbb{N}).
\ee
Equation (\ref{CFD-3}) means that the fractional derivatives 
$\mathbb{D}_C^{-} \left[ \alpha_j \atop j \right]$
of the odd orders $\alpha_j$ are local operators
represented by the usual derivatives of integer orders.
Note that the continuum derivative
$\mathbb{D}_C^{-} \left[ 2 m \atop j \right]$, 
where $m \in \mathbb{N}$,
cannot be considered as a usual local derivative 
${\partial^{2m}}/{\partial x^{2m}_j}$.
of the order $2m$ with respect to $x_j$.


\end{document}